\documentclass[smallextended]{svjour3}     
\smartqed  
\usepackage{graphicx}
\usepackage{multirow}
\begin{document}

\title{The Dark UNiverse Explorer (DUNE): Proposal to ESA's Cosmic Vision}


\author{A. Refregier and the DUNE collaboration\footnote{\normalsize{\bf Co-investigators:} A. Refregier ({\bf PI}, CEA Saclay) \and
 M. Douspis (IAS Orsay) \and Y. Mellier (IAP Paris) \and
 B. Milliard (LAM Marseille)
\and  P. Schneider (U. Bonn) \and H.-W. Rix (MPIA) \and
R. Bender (MPE Garching) \and F. Eisenhauer (MPE Garching)
% ...
}
}


\institute{A. Refregier \at
              Service d'Astrophysique\\ CEA Saclay, bat. 709\\ 91191 Gif sur Yvette, France\\
              \email{refregier@cea.fr}           
}

\date{Received: date / Accepted: date}

\maketitle

\begin{abstract}
The Dark UNiverse Explorer (DUNE) is a wide-field space imager whose
primary goal is the study of dark energy and dark matter with
unprecedented precision. For this purpose, DUNE is optimised for the
measurement of weak gravitational lensing but will also provide
complementary measurements of baryonic accoustic oscillations, cluster
counts and the Integrated Sachs Wolfe effect. Immediate auxiliary
goals concern the evolution of galaxies, to be studied with unequalled
statistical power, the detailed structure of the Milky Way and nearby
galaxies, and the demographics of Earth-mass planets. DUNE is an
Medium-class mission which makes use of readily available components,
heritage from other missions, and synergy with ground based facilities
to minimise cost and risks. The payload consists of a 1.2m telescope
with a combined visible/NIR field-of-view of 1 deg$^{\rm 2}$. DUNE
will carry out an all-sky survey, ranging from 550 to 1600nm, in one
visible and three NIR bands which will form a unique legacy for
astronomy.  DUNE will yield major advances in a broad range of fields
in astrophysics including fundamental cosmology, galaxy evolution, and
extrasolar planet search. DUNE was recently selected by ESA as one of
the mission concepts to be studied in its Cosmic Vision programme. 


\end{abstract}

\section{Introduction}      

Over the past decade, our large-scale view of the Universe has
undergone a revolution. Cosmologists have agreed on a standard model
that matches a wide range of astronomical data (eg. Spergel et
al. 2007). However, this $\Lambda$CDM concordance model relies on
three ingredients whose origin and nature are unknown: dark matter,
dark energy and fundamental fields driving a period of inflation,
during which density fluctuations are imprinted on the Universe. All
these elements of the model represent new aspects of fundamental
physics, which can best be studied via astronomy. The nature of the
dark energy, which now comprises the bulk of the mass-energy budget of
the Universe, will determine the ultimate fate of the Universe and is
among the deepest questions in physics.

The most powerful tool that can be brought to bear on these problems
is weak gravitational lensing of distant galaxies; this forms the core
of the DUNE mission\footnote{for further information on DUNE:
  www.dune-mission.net}. Gravitational deflection of light by
intervening dark matter concentrations causes the images of background
galaxies to acquire an additional ellipticity of order of a percent,
which is correlated over scales of tens of arcminutes. Measuring this
signature probes the expansion history in two complementary ways: (1)
geometrically, through the distance-redshift relation, and (2)
dynamically, through the growth rate of density fluctuations in the
Universe.

Utilisation of these cosmological probes relies on the measurement of
image shapes and redshifts for several billion galaxies. The
measurement of galaxy shapes for weak lensing imposes tight
requirements on the image quality which can only be met in the absence
of atmospheric phase errors and in the thermally stable environment of
space. For this number of galaxies, distances must be estimated using
photometric redshifts, involving photometry measurements over a wide
wavelength range in the visible and near-IR. The necessary visible
galaxy colour data can be obtained from the ground, using current or
upcoming instruments, complementing the unique image quality of space
for the measurement of image distortions. However, at wavelengths
beyond 1$\mu$m, we require a wide NIR survey to depths that are only
achievable from space.

Given the importance of the questions being addressed and to provide
systematic cross-checks, DUNE will also measure Baryon Acoustic
Oscillations, the Integrated Sachs-Wolfe effect, and galaxy Cluster
Counts. Combining these independent cosmological probes, DUNE will
tackle the following questions: What are the dynamics of dark energy?
What are the physical characteristics of the dark matter?  What are
the seeds of structure formation and how did structure grow?  Is
Einstein's theory of General Relativity the correct theory of gravity?

DUNE will combine its unique space-borne observation with existing and
planned ground-based surveys, and hence increases the science return
of the mission while limiting costs and risks. The panoramic visible
and NIR surveys required by DUNE's primary science goals will afford
unequalled sensitivity and survey area for the study of galaxy
evolution and its relationship with the distribution of the dark
matter, the discovery of high redshift objects, and of the physical
drivers of star formation. Additional surveys at low galactic
latitudes will provide a unique census of the Galactic plane and
earth-mass exoplanets at distances of 0.5-5 AU from their host star
using the microlensing technique. These DUNE surveys will provide a
unique all-sky map in the visible and NIR and thus complement other
space missions such as Planck, WMAP, eROSITA, JWST, and WISE. The
following describes the science objectives, instrument concept and
mission profile (see Table~\ref{table:summary} for a baseline
summary). A description of an earlier version of the mission without
NIR capability and developped during a CNES phase 0 study can be found
in Refregier et al 2006 and Grange et al. 2006.
\begin{table}
\caption{DUNE Baseline summary} 
\label{table:summary}
 \begin{tabular}{|l|l|}
\hline
Science objectives & Must: Cosmology and Dark Energy. Should: Galaxy formation\\
 & Could: Extra-solar planets\\
\hline
Surveys & Must: 20,000 deg$^2$ extragalactic, Should: Full sky (20,000
deg$^2$ \\
& Galactic), 100 deg$^2$ medium-deep. Could: 4 deg$^2$ planet hunting\\
\hline
Requirements & 1 visible band (R+I+J) for high-precision shape measurements,\\
 &  3 NIR bands (Y, J, H) for photometry\\
\hline
Payload & 1.2m telescope, Visible \& NIR cameras with 0.5 deg$^2$ FOV
each\\
\hline
Service module & Mars/Venus express, Gaia heritage \\
\hline
Spacecraft & 2013kg launch mass\\
\hline
Orbit & Geosynchronous\\
\hline
Launch & Soyuz S-T Fregat\\
\hline
Operations & 4 year mission\\
\hline
 \end{tabular}
\end{table}

\section{\label{section2}Science Objectives}          

The DUNE mission will investigate a broad range of astrophysics and
fundamental physics questions detailed below. Its aims are twofold:
first study dark energy and measure its equation of state parameter
$w$ (see definition below) and its evolution with a precision of 2\%
and 10\% respectively, using both expansion history and structure
growth, second explore the nature of dark matter by testing the Cold
Dark Matter (CDM) paradigm and by measuring precisely the sum of the
neutrino masses. At the same time, it will test the validity of
Einstein's theory of gravity. In addition, DUNE will investigate how
galaxies form, survey all Milky-Way-like galaxies in the 2$\pi$
extra-galactic sky out to $z \sim 2$ and detect thousands of galaxies
and AGN at $6<z<12$. It will provide a detailed visible/NIR map of
the Milky Way and nearby galaxies and provide a statistical
census of exoplanets with masses above 0.1 Earth mass and orbits
greater than 0.5 AU

\subsection{Understanding Dark Energy}

\begin{figure}
\begin{center}
\includegraphics[width=0.5\textwidth,angle=0]{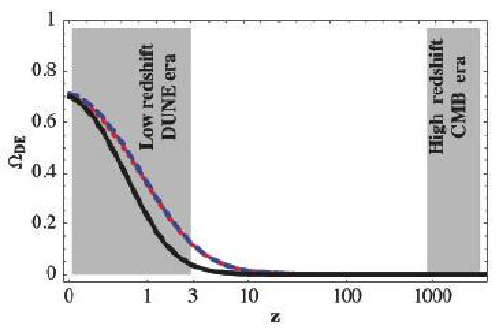}\includegraphics[width=0.5\textwidth,angle=0]{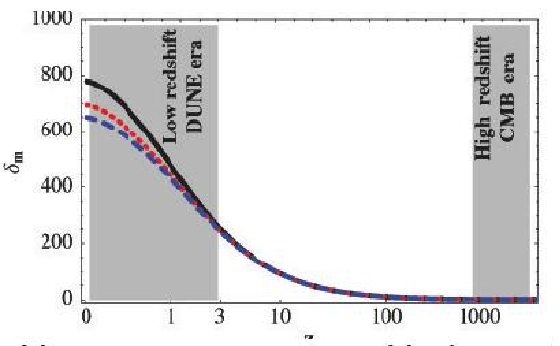}
\end{center}
\caption{Effects of dark energy. Left: Fraction of the density of the
Universe in the form of dark energy as a function of redshift $z$.,
for a model with a cosmological constant ($w=-1$, black solid line),
dark energy with a different equation of state ($w=-0.7$, red dotted
line), and a modified gravity model (blue dashed line). Dark energy
becomes dominant in the low redshift Universe era probed by DUNE,
while the early Universe is probed by the CMB. Right: Growth factor of
structures for the same models. Only by measuring the geometry
(bottom panel) and the growth of structure (top panel) at low
redshifts can a modification of dark energy be distinguished from that
of gravity. Weak lensing measures both effects.  }
\label{figc1}
\end{figure}

A variety of independent observations overwhelmingly indicate that the
cosmological expansion began to accelerate when the Universe was
around half of its present age. Presuming the correctness of general
relativity this requires a new energy component known as dark
energy. The simplest case would be Einstein's cosmological constant
($\Lambda$), in which the dark energy density would be exactly
homogeneous and independent of time. However, the description of
vacuum energy from current Particle Physics concepts conflicts by 120
orders of magnitude with the observed value, and the discrepancy is
still not understood. Cosmologists are thus strongly motivated to
consider models of a dynamical dark energy, or even to contemplate
modifications to General Relativity.  Explaining dark energy may well
require a radical change in our understanding of Quantum Theory or
Gravity, or both. One of the major aims of DUNE is to determine
empirically which of these alternatives is to be preferred.  The
properties of dark energy can be quantified by considering its
equation of state parameter $w=p/\rho c^2$, where $p$ and $\rho$ are
its effective pressure and density. Unlike matter, dark energy has the
remarkable property of having negative pressure ($w<0$) and thus of
driving the Universe into a period of accelerated expansion (if
$w<-1/3$). The latter began relatively recently, around $z \le 1$. If
the dark energy resembles a cosmological constant ($w=-1$), it can
only be directly probed in the low-redshift Universe (see
Fig.~\ref{figc1}). This expansion history can be measured in two
distinct ways (see Fig.~\ref{figc1}): (1) the distance-redshift
relation $D(z)$; (2) the growth of structure (i.e. galaxies and
clusters of galaxies). The $D(z)$ relation can be probed geometrically
using 'standard candles' such as supernovae, or via measures of the
angular diameter distance from gravitational lensing or from the
``standard rod'' of Baryon Acoustic Oscillations (BAO). The
accelerated expansion slows down the gravitational growth of density
fluctuations; this growth of structure can be probed by comparing the
amplitude of structure today relative to that when the CMB was formed.
Many models for dark energy and modifications to gravity have been
proposed in which the equation of state parameter $w$ vary with
time. A convenient approximation is a linear dependence on the scale
factor $a=1/(1+z)$: $w(a)=w_n+(a_n-a)w_a$, where $w_n$ is the value of
the equation of state at a pivot scale factor $a_n$ (close to 0.6 for
most probes) and $w_a$ describes the redshift evolution. The goal of
future surveys is to measure $w_n$ and $w_a$ to high precision. To
judge their relative strengths we use a standard dark energy figure of
merit (FoM) (Albrecht et al. 2006), which we define throughout this
proposal as: $FoM=1/(\Delta w_n \Delta w_a)$, where $\Delta w_n$ and
$\Delta w_a$ are the (1$\sigma$) errors on the equation of state
parameters. This FoM is inversely proportional to the area of the
error ellipse in the ($w_n-w_a$) plane.

\subsection{DUNE's Cosmological Probes}
\begin{figure}
\begin{center}
\includegraphics[width=0.5\textwidth,angle=0]{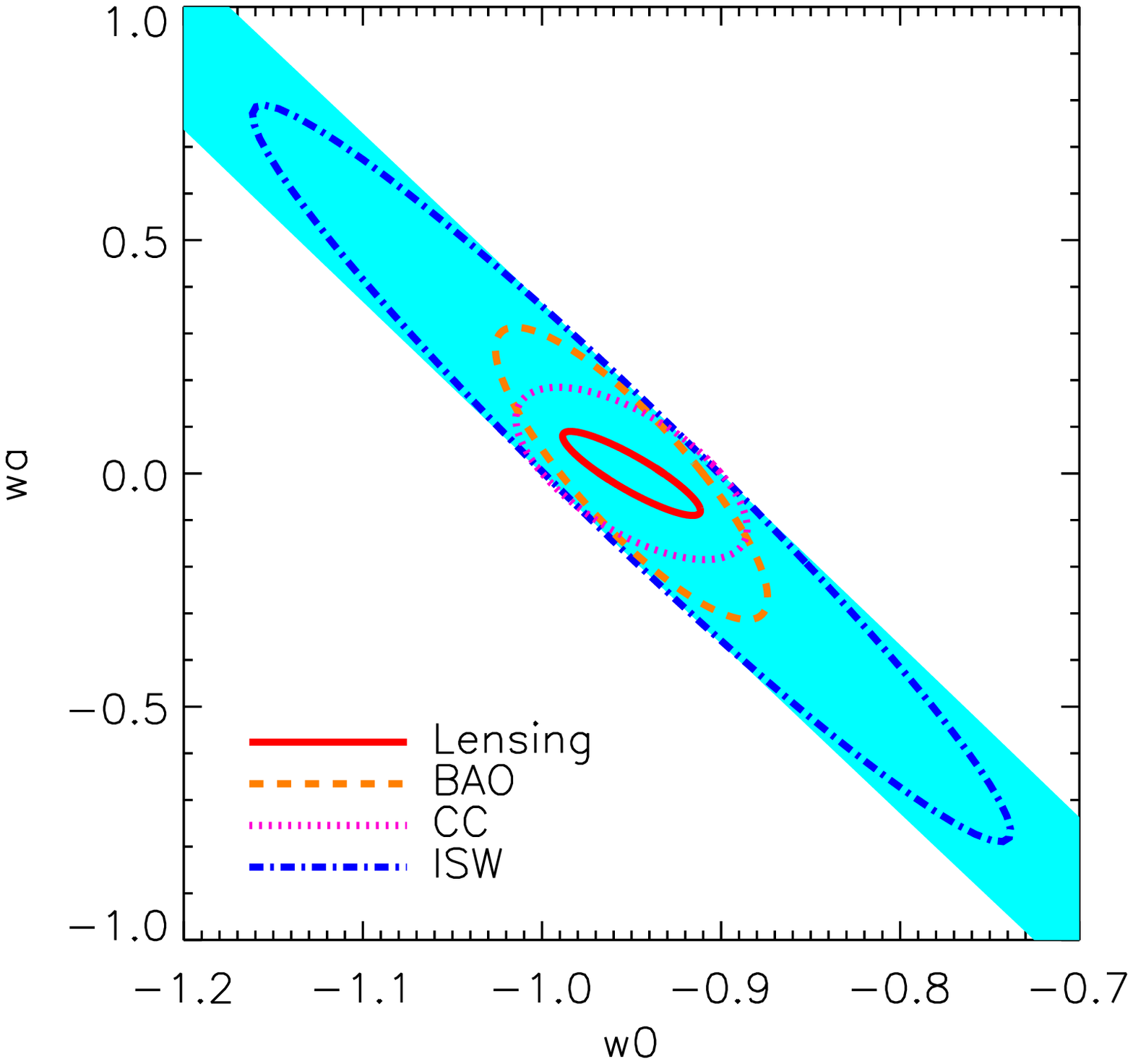}\includegraphics[width=0.5\textwidth,angle=0]{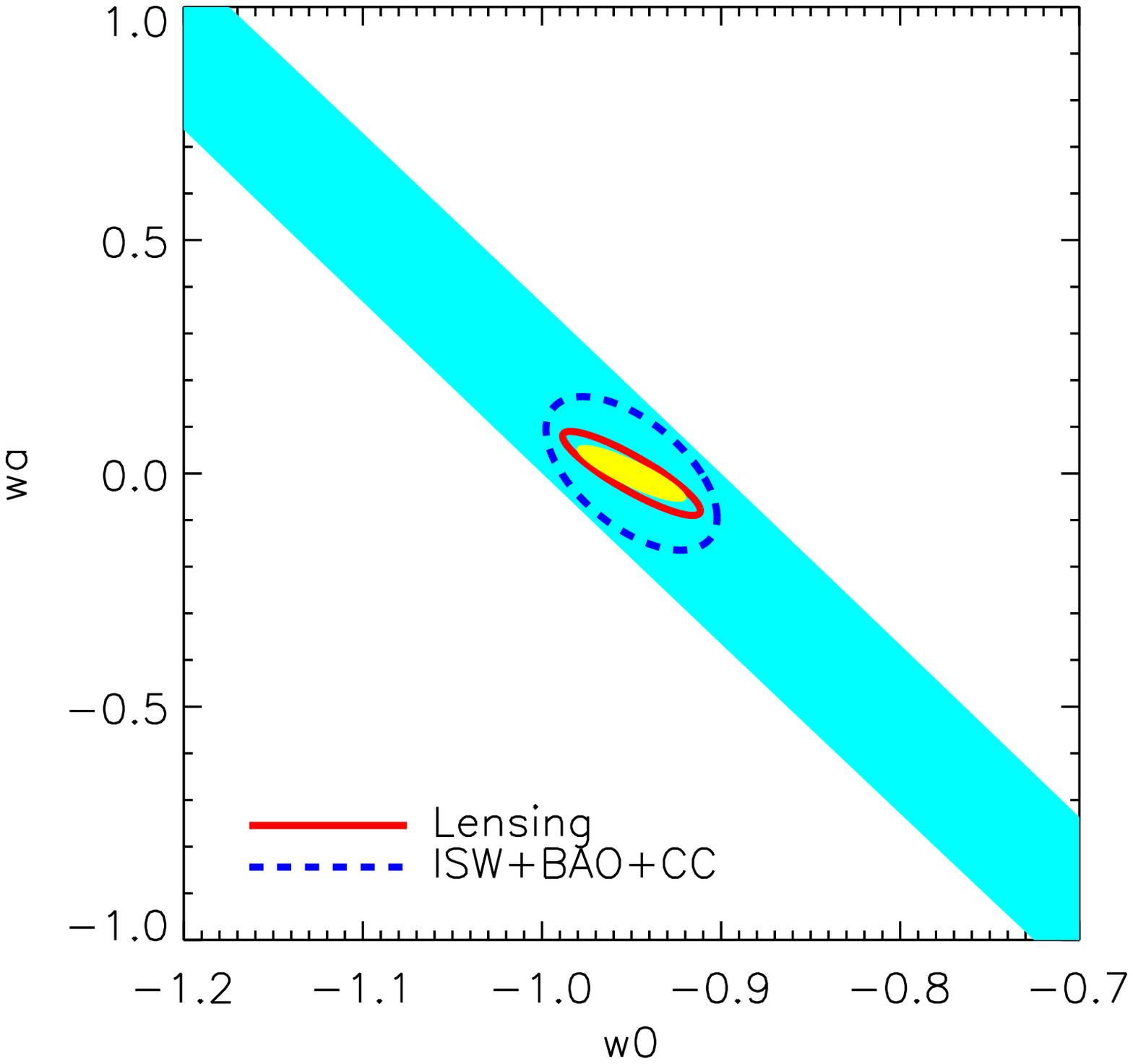}
\end{center}
\caption{Left: Expected errors on the dark energy equation of state
  parameters for the four probes used by DUNE (68\% statistical
  errors). The light blue band indicates the expected errors from
  Planck. Of the four methods, weak lensing clearly has the greatest
  potential. Right: The combination of BAO, CC and ISW (red solid
  line) begins to reach the potential of the lensing survey (blue
  dashed line) and provides an additional cross-check on
  systematics. The yellow ellipse corresponds to the combination of
  all probes and reaches a precision on dark energy of 2\% on $w_n$
  and 10\% on $w_a$.}
\label{figc3}
\end{figure}

DUNE will deduce the expansion history from the two methods,
distance-redshift relation and growth of structure. DUNE has thus the
advantage of probing the parameters of dark energy in two independent
ways.  A single accurate technique can rule out many of the suggested
members of the family of dark energy models, but it cannot test the
fundamental assumptions about gravity theory. If General Relativity is
correct, then either $D(z)$ or the growth of structure can determine
the expansion history. In more radical models that violate General
Relativity, however, this equivalence between $D(z)$ and growth of
structure does not apply (see Fig.~\ref{figc1}).  For this purpose,
DUNE will use a combination of the following cosmological probes. The
precision on Dark Energy parameters achieved by DUNE's weak lensing
survey and complementary probes described below is shown in
Fig.~\ref{figc3} and Table~\ref{tableC2}.

{\it Weak Lensing - A Dark Universe Probe:} 
As light from galaxies travels towards us, its path is deflected by
the intervening mass density distribution, causing the shapes of these
galaxies to appear distorted by a few percent. The weak lensing method
measures this distortion by
correlating the shapes of background galaxies to probe the density
field of the Universe. By dividing galaxies into redshift (or
distance) bins, we can examine the growth of structure and make
three-dimensional maps of the dark matter. An accurate lensing survey,
therefore, requires precise measurements of the shapes of galaxies as
well as information about their redshifts. High-resolution images of
large portions of the sky are required, with low levels of systematic
errors that can only be achieved via observations from a thermally
stable satellite in space. Analyses of the dark energy require precise
measurements of both the cosmic expansion history and the growth of
structure. Weak lensing stands apart from all other available methods
because it is able to make accurate measurements of both
effects. Given this, the optimal dark energy mission (and dark sector
mission) is one that is centred on weak gravitational lensing and is
complemented by other dark energy probes.

{\it Baryon Acoustic Oscillations (BAO) -- An Expansion History
  Probe:} 
The scale of the acoustic oscillations caused by the
coupling between radiation and baryons in the early Universe can be
used as a 'standard ruler' to determine the distance-redshift
relation. Using DUNE, we can perform BAO measurements using
photometric redshifts yielding the three-dimensional positions of a
large sample of galaxies. All-sky coverage in the NIR enabled by DUNE,
impossible from the ground, is crucial to reach the necessary
photometric redshift accuracy for this BAO survey.

{\it Cluster Counts (CC) -- A Growth of Structure Probe:} 
Counts of the abundance of galaxy clusters (the most massive bound
objects in the Universe) as a function of redshift are a powerful
probe of the growth of structure. There are three ways to exploit the
DUNE large-area survey, optimised for weak lensing, for cluster
detection: strong lensing; weak lensing; and optical richness.

{\it Integrated Sachs-Wolfe (ISW) Effect -- A Higher Redshift
  Probe:} The ISW effect is the change in CMB photon energy as it
passes through a changing potential well.  Its presence indicates
either space curvature, a dark energy component or a modification to
gravity. The ISW effect is measured by cross-correlating the CMB with
a foreground density field covering the entire extra-galactic sky, as
measured by DUNE. Because it is a local probe of structure growth, ISW
will place complementary constraints on dark energy, at higher
redshifts, relative to the other probes (Douspis et al. 2008).

\subsection{Understanding Dark Matter}

Besides dark energy, one major component of the concordance model of
cosmology is dark matter ($\sim90$\% of the matter in the Universe,
and $\sim 25$\% of the total energy). The standard assumption is that
the dark matter particle(s) is cold and non-collisional (CDM). Besides
direct and indirect dark matter detection experiments, its nature may
well be revealed by experiments such as the Large Hadron Collider
(LHC) at CERN, but its physical properties may prove to be harder to
pin down without astronomical input.  one way of testing this is to
study the amount of substructure in dark matter halos on scales
1-100'', which can be done using high order galaxy shape measurements
and strong lensing with DUNE. Weak lensing measurements can constrain
the total neutrino mass and number of neutrino species through
observations of damping of the matter power spectrum on small
scales. Combining DUNE measurements with Planck data would reduce the
uncertainty on the sum of neutrino masses to 0.04eV, and may therefore
make the first measurement of the neutrino mass (Kitching et al 2008).

\subsection{Understanding the Seeds of Structure Formation}

It is widely believed that cosmic structures originated from vacuum
fluctuations in primordial quantum fields stretched to cosmic scales
in a brief period during inflation. In the most basic inflationary
models, the power spectrum of these fluctuations is predicted to be
close to scale-invariant, with a spectral index $n_s$ and amplitude
parameterised by $\sigma_8$. As the Universe evolved, these initial
fluctuations grew.  CMB measurements probe their imprint on the
radiation density at $z \sim 1100$. Density fluctuations continued to
grow into the structures we see today. Weak lensing observations with
DUNE will lead to a factor of 20 improvement on the initial
conditions as compared to CMB alone (see Table~\ref{tableC2}).

\subsection{Understanding Einstein's Gravity}
Einstein's General Theory of Relativity, the currently accepted theory
of gravity, has been well tested on solar system and galactic
scales. Various modifications to gravity on large scales (e.g. by
extra dimensions, superstrings, non-minimal couplings or additional
fields) have been suggested to avoid dark matter and dark energy. The
weak lensing measurements of DUNE will be used to test the underlying
theory of gravity, using the fact that modified gravity theories
typically alter the relation between geometrical measures and the
growth of structure (see Fig.~\ref{figc1}). DUNE can measure the the
growth factor exponent $\gamma$ with a precision of 2\%.

\begin{table}
\caption{Dark energy and initial conditions figures of merit for DUNE
  and Planck.} 
\label{tableC2}
 \begin{tabular}{|l|r|r|r|r|r|r|}
\hline
& \multicolumn{2}{c}{Dark Energy Sector}\vline & \multicolumn{2}{c}{Initial Conditions
Sector}\vline& DE &IC \\ \hline
 & $\Delta w_n$ & $\Delta w_a$ & $\Delta \sigma_8$ & $\Delta n_s$ & FoM & FoM \\ \hline
Planck &0.03 &3  &0.03  &0.004&11  &8,000 \\ \hline
DUNE   &0.02 &0.1&0.006 &0.01 &500 &17,000 \\ \hline
DUNE + Planck& 0.01&0.05&0.002&0.003&2000&170,000\\ \hline
 \multicolumn{5}{l}{Factor improvement of DUNE + Planck over Planck only}&180&20 \\ \hline
 \end{tabular}
\end{table}
\par\bigskip
Meeting the above cosmological objectives necessitates an
extra-galactic all-sky survey (DASS-EX) in the visible/NIR with
galaxies at a median redhift of $z \sim 1$. To this survey, will be
added a shallower survey of the Galactic plane (DASS-G) which will
complete the coverage to the full sky, as well as a medium-deep survey
of 100 deg$^{\rm 2}$ (DUNE-MD) and a pencil beam microlensing survey
for planets in the Galactic bulge.

Focussed on the dark sector, DUNE will produce an invaluable broad survey
legacy.  DASS will cover a 10000 times larger area than other
optical/near-IR surveys of the same or better resolution, will be 4mag
deeper than the GAIA photometry and six times higher resolution than
the SDSS. In the infrared, DASS-EX will be nearly 1000 times deeper
(in J) than the all-sky 2MASS survey with an effective
search volume
which will be 5000-fold that of the UKIDDS large area survey currently
underway, and 500-fold that of the proposed VISTA Hemisphere
Survey. It would take VISTA 8000 years to match DASS-EX depth and
20,000 deg$^{\rm 2}$ area coverage. DASS-MD will bridge the gap
between DASS-EX and expected JWST surveys.

\subsection{Tracking the Formation of Galaxies and AGN with DUNE}

While much progress has been made in understanding the formation of
large scale structure, there are still many problems in forming
galaxies within this structure with the observed luminosity function
and morphological properties. This is now a major problem in
astronomy. Obtaining deep high spatial resolution near-IR images will
be central to the study galaxy morphology and clustering. A large area
survey is required for rare but important events, such as the merger
rate of very massive galaxies. DUNE will deliver this key capability.

Using DUNE's weak lensing maps, we will study the relationship between
galaxy mass and light, the bias, by mapping the total mass density and
the stellar mass and luminosity.
Galaxy clusters are the largest scale signposts of
structure formation.
While at present only a few massive clusters
at $z>1$ are known, DUNE will find hundreds of Virgo-cluster-mass
objects at $z>2$, and several thousand clusters of M=$1-2 \times
10^{13}$M$_{\odot}$. The latter are the likely environments in which the peak
of QSO activity at $z\sim2$ takes place, and hold the empirical
key to understanding the heyday of QSO activity.

Using the Lyman-dropout technique in the near-IR, the DUNE-MD survey
will be able to detect the most luminous objects in the early Universe
($z>6$): $\sim 10^4$ star-forming galaxies at $z\sim8$ and up to
$10^3$ at $z\sim10$, for SFRs $>30-100$M$_{\odot}$/yr. It will also be able
to detect significant numbers of high-$z$ quasars: up to $10^4$ at
$z\sim7$, and $10^3$ at $z\sim9$. These will be central to understanding the
reionisation history of the Universe.

Dune will also detect a very large number of strong lensing systems: 
about $10^5$ galaxy-galaxy lenses, $10^3$ galaxy-quasar lenses and
5000 strong lensing arcs in clusters (see Menegetthi et al. 2007).  It
is also estimated that several tens of galaxy-galaxy lenses will be
\emph{double} Einstein rings (Gavazzi et al. 2008), which are powerful
probes of the cosmological model as they simultaneously probe several redshifts.

In addition, during the course of the DUNE-MD survey (over 6 months),
we expect to detect $\sim 3000$ Type Ia Supernovae with redshifts up
to $z\sim0.6$ and a comparable number of Core Collapse SNe (Types II
and Ib/c) out to $z\sim0.3$. This will lead to measurement of SN rates
thus providing information on their progenitors and on the star
formation history.

\subsection{Studying the Milky Way with DUNE}

DUNE is also primed for a breakthrough in Galactic astronomy. DASS-EX,
complemented by the shallower survey of the Galactic plane (with
$|b|<30\; deg$) will provide all-sky high resolution (0.23'' in the wide red
band, and 0.4'' in YJH) deep imaging of the stellar content of the
Galaxy, allowing the deepest detailed structural studies of the thin
and thick disk components, the bulge/bar, and the Galactic halo
(including halo stars in nearby galaxies such as M31 and M33) in bands
which are relatively insensitive to dust in the Milky Way.

DUNE will be little affected by extinction and will supersede by
orders of magnitude all of the ongoing surveys in terms of angular
resolution and sensitivity.
DUNE will thus
enable the most comprehensive stellar census of late-type dwarfs and
giants, brown dwarfs, He-rich white dwarfs, along with detailed
structural studies, tidal streams and merger fragments.  DUNE's
sensitivity will also open up a new discovery space for rare stellar
and low-temperature objects via its H-band imaging.  Currently, much
of Galactic structure studies are focussed on the halo. Studying the
Galactic disk components requires the combination of spatial
resolution (crowding) and dust-penetration (H-band) that DUNE can
deliver.

Beyond our Milky Way, DUNE will also yield the most detailed and
sensitive survey of structure and substructure in nearby galaxies,
especially of their outer boundaries, thus constraining their merger
and accretion histories.

\subsection{Search for Exo-Planets}
\begin{figure}
\begin{center}
\includegraphics[width=7cm, height=6cm, angle=0]{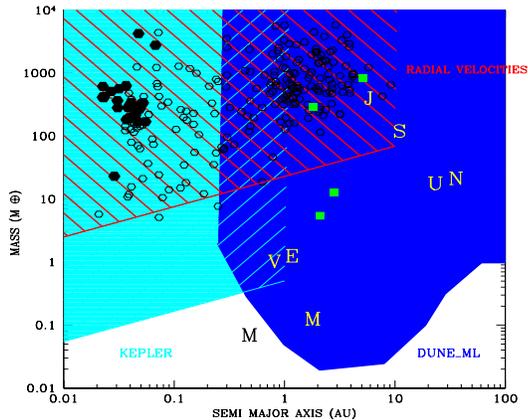}
\caption{Exoplanet discovery parameter space (planet mass vs orbit size)
showing for reference the 8 planets from our solar system (labeled as letters),
those detected by Doppler wobble (T), transit (circle), and
microlensing. We outline regions that can be probed by different
methods. Note the uniqueness of the parameter space probed by DUNE
compared to other techniques. 
}
\label{figc5}
\end{center}
\end{figure}

The discovery of extrasolar planets is the most exciting development
in astrophysics over the past decade, rivalled only by the discovery
of the acceleration of the Universe. Space observations (e.g. COROT, KEPLER), supported by
ground-based high-precision radial velocity surveys will probe
low-mass planets (down to $1 M_\oplus$). DUNE is also
perfectly suited to trace the distribution of matter on very small
scales those of the normally invisible extrasolar planets.  Using
microlensing effect, DUNE can provide a statistical census of
exoplanets in the Galaxy with masses over $0.1 M_\oplus$ from orbits
of 0.5 AU to free-floating objects. This includes analogues to all the
solar system's planets except for Mercury, as well as most planets
predicted by planet formation theory.  Microlensing is the temporary
magnification of a Galactic bulge source star by the gravitational
potential of an intervening lens star passing near the line of
sight. A planet orbiting the lens star, will have an altered
magnification, showing a brief flash or a dip in the observed light
curve of the star(see Fig. \ref{figc5}). 
Because of atmospheric seeing (limiting the monitoring to large source
stars), and poor duty cycle even using networks, ground-based
microlensing surveys are only able to detect a few to 15 $M_\oplus$
planets in the vicinity of the Einstein ring radius (2-3 AU). The high
angular resolution of DUNE, and the uninterrupted visibility and NIR
sensitivity afforded by space observations will provide detections of
microlensing events using as sources G and K bulge dwarfs stars and
therefore can detect planets down to $0.1-1 M_\odot$ from orbits of
0.5 AU.  Moreover, there will be a very large number of transiting hot
Jupiters detected towards the Galactic bulge as 'free' ancillary
science. A space-based microlensing survey is thus the only way to
gain a comprehensive census and understanding of the nature of
planetary systems and their host stars. We also underline that the
planet search scales linearly with the surface of the focal plane and
the duration of the experiment.

\section{DUNE Surveys: the Need for All-Sky Imaging from Space}
\begin{figure}
 \includegraphics[width=0.5\textwidth,height=5cm,angle=0]{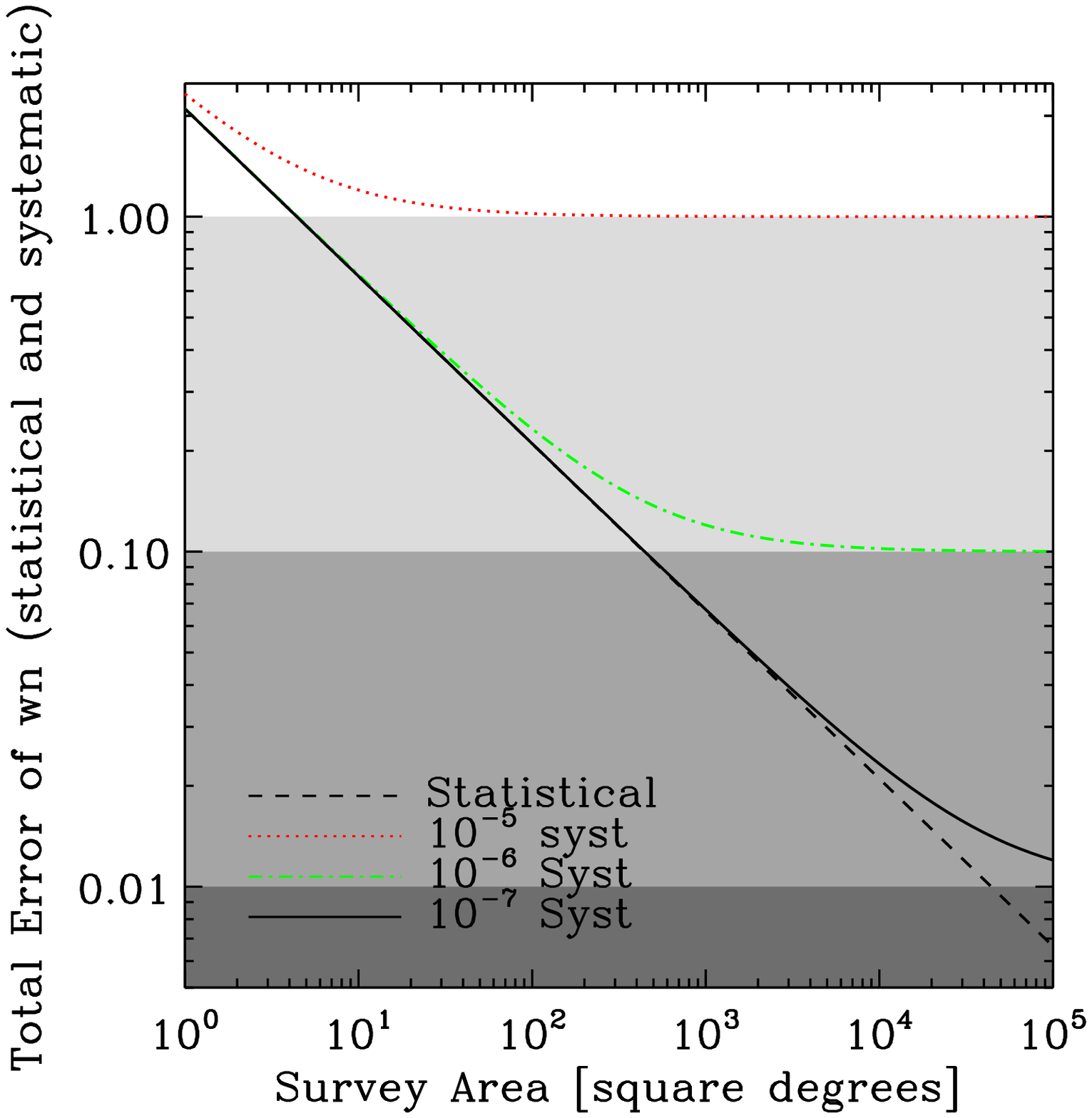}\includegraphics[width=0.5\textwidth,angle=0]{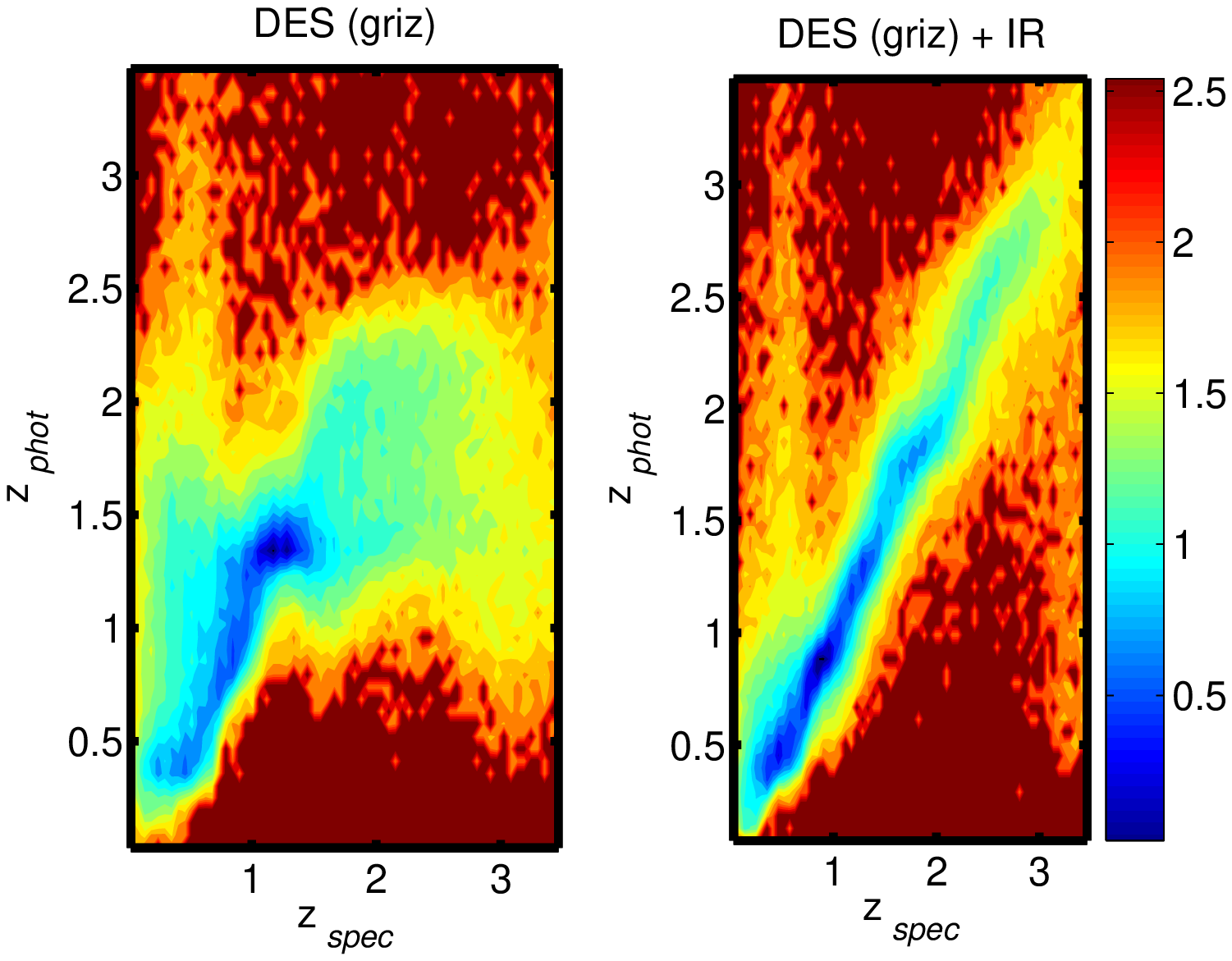}
\caption{Left: Error on the dark energy equation of state
parameter $w_n$ as a function of weak lensing survey area (in deg$^{\rm
2}$) for several shape measurement systematic levels (assuming 40
galaxies/amin$^2$ with a median redshift $z_m$=1). An area of 20,000 deg$^2$
and a residual systematic shear variance of
$\sigma_{sys}^2<10^{-7}$ is required to achieve the DUNE objective
(error on $w_n$ better than 2\%). 
Right (from Abdalla et
al. 2007): Photometric redshift performance for a DES-like ground survey
with and without the DUNE NIR bands (J,H). The deep NIR photometry,
only achievable in space, results in a dramatic reduction of the
photometric redshift errors and catastrophic failures which are needed for all
the probes (weak lensing, BAO, CC, ISW).}
\label{fig:req}
\end{figure}

There are two key elements to a high precision weak lensing survey: a
large survey area to provide large statistics, and the control of
systematic errors. Figure \ref{fig:req} shows that to
reach our dark energy target (2\% error on $w_n$) a survey of 20,000
square degrees with galaxies at $z\sim1$ is required. This result is
based on detailed studies showing that, for a fixed observing time,
the accuracy of all the cosmological parameters is highest for a wide
rather than deep survey (Amara \& Refregier 2007a, 2007b).  This required
survey area drives the choice of a 1.2m telescope and a combined
visible/NIR FOV of 1 deg$^{\rm 2}$ for the DUNE baseline.

Ground based facilities plan to increase area coverage, but they will
eventually be limited by systematics inherent in ground based
observations (atmospheric seeing which smears the image, instabilities
of ground based PSFs, telescope flexure and wind-shake, and
inhomogeneous photometric calibrations arising from seeing
fluctuations). The most recent ground-based wide field imagers
(e.g. MegaCam on CFHT, and Subaru) have a stochastic variation of the
PSF ellipticity of the order of a few percent, i.e. of the same order
of magnitude as the sought-after weak lensing signal.  Current
measurements have a residual shear systematics variance of
$\sigma_{sys}^2 \sim 10^{-5}$, as indicated with both the results of
the STEPII program and the scatter in measured value of
$\sigma_8$. This level of systematics is comparable to the statistical
errors for surveys that cover a few tens of square degree
(Fig. \ref{fig:req}). As seen on the figure, to reach DUNE's dark
energy targets, the systematics must be at the level of
$\sigma_{sys}^2 \sim 10^{-7}$, 100 times better than the current level
(see Amara \& Refregier 2007b for details). While ground based surveys
may improve their systematics control, reaching this level will be an
extreme challenge. One ultimate limit arises from the finite
information contained in the stars used to calibrate the PSF, due to
noise and pixelisation. Simulations by Paulin-Henriksson et al. (2008)
show that, to reach our systematic targets, the PSF must remain
constant (within a tolerance of 0.1\%) over 50 arcmin$^2$ (which
corresponds to $\sim 50$ stars). While this is prohibitive from the
ground, we have demonstrated during a CNES phase 0 study (Refregier et
al. 2006), that with the careful design of the instrument, this can be
achieved from space.  In addition to shape measurements, wide area
imagining surveys use photometric information to measure the redshift
of galaxies in the images. Accurate measurements of the photometric
redshifts require the addition of NIR photometry (an example of this
is shown in Fig. \ref{fig:req}, right panel, and also Abdalla et
al. 2007). Such depths in the NIR cannot be achieved from the ground
over wide area surveys and can only be done from space.

\par\bigskip

To achieve the scientific goals listed in section \ref{section2}, DUNE will
perform four surveys detailed in the following and in Table \ref{tableC5}. 

\subsection{Wide Extragalactic Survey: DASS-EX }
To measure dark energy to the required precision, we need to make
measurements over the entire extra-galactic sky to a depth which
yields 40 gal/arcmin$^2$ useful for lensing with a median redshift
$z_m \simeq 0.9$. This can be achieved with a survey (DASS-EX) that
has AB-magnitude limit of 24.5 (10$\sigma$ extended source) in a broad
red visible filter (R+I+Z). Based on the fact that DUNE focuses on
observations that cannot be obtained from the ground, the wide survey
relies on two unique factors that are enabled by space: image quality
in the visible and NIR photometry.  Central to shape measurements for
weak lensing the PSF of DUNE needs to be sampled better than 2-2.5
pixels per FWHM (Paulin-Henriksson et al. 2008), to be constant over
50 stars around each galaxy (within a tolerance of $\sim 0.1\%$ in
shape parameters), and to have a wavelength dependence which can be
calibrated.  Accurate measurement of the redshift of distant galaxies
($z \sim 1$) requires photometry in the NIR where galaxies have a
distinctive feature (the 4000$\AA$ break). Deep NIR photometry
requires space observations. The bands Y, J and H are the perfect
synergy for ground based survey complement (see Abdalla et al. 2007
for discussion), as recommended by the ESO/ESA Working Group on
Fundamental Cosmology (Peacock et al. 2006).

\subsection{Legacy surveys: DASS-G, DUNE-MD, and DUNE-ML}

We propose to allocate six months to a medium deep survey (DUNE-MD) with
an area of 100 deg$^2$ to magnitudes of 26 in Y, J and H, located at
the N and S ecliptic poles. This survey can be used to calibrate DUNE
during the mission, by constructing it from a stack of $>30$
sub-images too achieve the required depths. DUNE will also perform a
wide Galactic survey (DASS-G) that will complement the 4$\pi$ coverage
of the sky and a microlensing survey (DUNE-ML). Both surveys require
short exposures. Together with the DASS-EX, these surveys need good
image quality with low level of stray light. A summary of all the
surveys is shown in Table \ref{tableC5}.

\begin{table}
\caption{Requirements and geometry for the four DUNE surveys.}
\label{tableC5}
 \begin{tabular}{|c|c|c|c|}
\hline
\multicolumn{4}{|c|}{\textbf{Wide Extragalactic Survey DASS-EX (must)}}\\ \hline
\multicolumn{2}{|c|}{Area}&\multicolumn{2}{|c|}{20,000 sq degrees  -- $|b|> 30 \deg$
}\\ \hline
\multirow{2}{*}{Survey Strategy}& Contiguous patches  & \multicolumn{2}{|c|}{$> 20 \deg \times 20 \deg$} \\  \cline{2-4}
                            & Overlap             & \multicolumn{2}{|c|}{$ 10 \%$} \\  \hline
\multicolumn{2}{|c|}{Shape Measurement Channel}& R+I+Z (550-920nm) & R+I+$Z_{AB}$ $<$24.5 (10$\sigma$ ext) \\ \hline
\multicolumn{2}{|c|}{                   }  & Y (920-1146nm)  & $Y_{AB}<$24 (5$\sigma$ point) \\ \cline{3-4}
\multicolumn{2}{|c|}{Photometric Channel}  & J (1146-1372nm) & $J_{AB}<$24 (5$\sigma$ point) \\ \cline{3-4}
\multicolumn{2}{|c|}{                   }  & H (1372-1600nm) & $H_{AB}<$24 (5$\sigma$ point) \\ \hline
\multirow{2}{*}{PSF} & Size  \& Sample      & 0.23" FWHM &  $>$ 2.2 pixels per FWHM \\\cline{2-4} 
                     & Stability      & \multicolumn{2}{|c|}{within tolerance of 50 stars}  \\ \hline
\multirow{2}{*}{Image Quality} & Dead pixels  &\multicolumn{2}{|c|}{$<$ 5 \% of final image}\\ \cline{2-4}
                             &  Linearity     &\multicolumn{2}{|c|}{Instrument calibratable for $1<$S/N$<1000$}\\ \hline                           
\multicolumn{4}{|c|}{\textbf{Medium Deep Survey DUNE-MD (should)}}\\ \hline  
\multicolumn{2}{|c|}{Area}&\multicolumn{2}{|c|}{ $\sim$100 sq degrees
--  Ecliptic poles}\\ \hline
Survey Strategy& Contiguous patches  & \multicolumn{2}{|c|}{Two patches each $7 \deg \times 7 \deg$} \\  \hline
\multicolumn{2}{|c|}{Photometric Channel}  & \multicolumn{2}{|c|}{  $Y_{AB}, \; J_{AB}, \; H_{AB} <$26 (5$\sigma$ point) -- for stack}\\ \hline
\multicolumn{2}{|c|}{PSF}            & \multicolumn{2}{c|}{Same conditions as the wide survey} \\ \hline
\multicolumn{4}{|c|}{\textbf{Wide Galactic Survey DASS-G (should)}}\\ \hline  
\multicolumn{2}{|c|}{Area}&\multicolumn{2}{|c|}{ 20,000 sq degrees  --
 $|b| < 30 \deg$}\\ \hline
\multicolumn{2}{|c|}{Shape Measurement Channel}&\multicolumn{2}{|c|}{$R+I+Z_{AB}<23.8$ ($5\sigma$ ext)}\\ \hline
\multicolumn{2}{|c|}{Photometric Channel}  & \multicolumn{2}{|c|}{  $Y_{AB}, \; J_{AB}, \; H_{AB} <$22 (5$\sigma$ point)}\\ \hline
PSF           &  Size        & \multicolumn{2}{|c|}{$< 0.3"$ FWHM}\\ \hline
\multicolumn{4}{|c|}{\textbf{Microlensing Survey DUNE-ML (could)}}\\ \hline  
\multicolumn{2}{|c|}{Area}&\multicolumn{2}{|c|}{ 4 sq degrees -- Galactic bulge}\\ \hline
Survey Strategy                       & Time sampling      & \multicolumn{2}{|c|}{Every 20 min -- 1 month blocks -- total of 3 months}\\ \hline
\multicolumn{2}{|c|}{Photometric Channel}  & \multicolumn{2}{|c|}{
  $Y_{AB}, \; J_{AB}, \; H_{AB} <$22 (5$\sigma$ point) -- per visit}\\ \hline
PSF &  Size        & \multicolumn{2}{|c|}{$< 0.4"$ FWHM}\\ \cline{2-4}
   \hline
\end{tabular}

\end{table}

\section{Mission Profile and Payload instrument} 
The mission design of DUNE is driven by the need for the stability of
the PSF and large sky coverage. PSF stability puts stringent
requirements on pointing and thermal stability during the observation
time. The 20,000 square degrees of DASS-EX demands high operational
efficiency, which can be achieved using a drift scanning mode (or Time
Delay Integration, TDI, mode) for the CCDs in the visible focal
plane. TDI mode necessitates the use of a counter-scanning mirror to
stabilize the image in the NIR focal plane channel.

The baseline for DUNE is a Geosynchronous Earth orbit (GEO), with a
low inclination and altitude close to a standard geostationary
orbit. Based on Phase 0 CNES study, this solution was chosen to meet
both the high science telemetry needs and the spacecraft low
perturbation requirements.  This orbit also provides substantial
launch flexibility, and simplifies the ground segment.

As for the PSF size and sampling requirements, a
baseline figure for the line-of-sight stability is 0.5 pixel (smearing
MTF $> 0.99$ at cut-off frequency), with the stability budget to be
shared between the telescope thermal stability (0.35 pixel) and the
attitude orbit control system (AOCS) (0.35 pixel). This implies a
line-of-sight stability better than 0.2 $\mu$rad over 375 sec (the
integration time across one CCD). This stringent requirement calls for
a minimalization of external perturbations which mainly consist of solar radiation
pressure and gravity gradient torques. A gravitational torque of 20
$\mu$Nm is acceptable, and requires an orbit altitude of at least
25,000 km. The attitude and orbit control design is based on proportional
actuators. 

A stable thermal environment is requested for the payload ($\sim 10
mK$ variation over 375sec), hence mission design requires a permanent
cold face for the focal plane radiators and an orbit that minimizes
heat load from the Earth. This
could be achieved by having the whole payload in a regulated temperature
cavity. 

A primary driver for the GEO orbit choice is the high data rate -- the
orbit must be close enough to Earth to facilitate the transmission of
the high amount of data produced by the payload every day (about
1.5~Tbits) given existing ground network facilities, while minimizing
communication downlink time during which science observations cannot
be made (with a fixed antenna).

The effects of the radiation environment at GEO, for both CCD bulk
damage induced by solar protons and false detections induced by
electrons with realistic shielding, is considered acceptable. However,
DUNE specific radiation tests on CCD sensors will be required as an
early development for confirming the measurement robustness to proton
induced damage. A High Elliptical Orbit (HEO) operating beyond the
radiation belt is an alternative in case electron radiation or thermal
constraints prevent the use of GEO.

\begin{figure}
\begin{center}
\includegraphics[width=0.75\textwidth,angle=0]{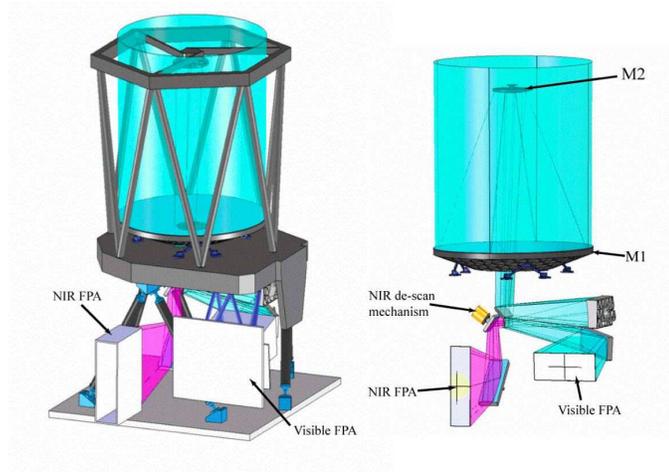}
\caption{Overview of all payload elements. }
\label{fig:4.1}
\end{center}
\end{figure}

The payload for DUNE is a passively cooled 1.2m diameter Korsch-like
three-mirror telescope with two focal planes, visible and NIR covering
1 square degree.  Figure~\ref{fig:4.1} provides an overview of the
payload.  The Payload module design uses Silicon Carbide (SiC)
technology for the telescope optics and structure. This provides low
mass, high stability, low sensitivity to radiation and the ability to
operate the entire instrument at cold temperature, typically below 170
K, which will be important for cooling the large focal planes. The two
FPAs, together with their passive cooling structures are isostatically
mounted on the M1 baseplate. Also part of the payload are the de-scan
mirror mechanism for the NIR channel and the additional payload data
handling unit (PDHU).

\subsection{Telescope}

The telescope is a Korsch-like $f/20$ three-mirror telescope. After
the first two mirrors, the optical bundle is folded just after passing
the primary mirror (M1) to reach the off-axis tertiary mirror. A
dichroic element located near the exit pupil of the system provides
the spectral separation of the visible and NIR channels. For the NIR,
the de-scan mechanism close to the dichroic filter allows for a
largely symmetric configuration of both spectral channels. The whole
instrument fits within a cylinder of 1.8m diameter and
2.65m length. The payload mass is approximately 500~kg, with 20\%
margin, and average/peak power estimates are 250/545~W.

Simulations have shown that the overall wavefront error (WFE) can be
contained within 50 nm r.m.s, compatible with the required
resolution. Distortion is limited to between 3-4$\mu$m, introducing an
0.15$\mu$rad fixed (hence accessible to calibration) displacement in
the object space.  The need to have a calibration of the PSF shape
error better than 0.1 \% over 50 arcmin$^2$ leads to a thermal
stability of $\sim 10$ mK over 375 s. Slow variations of solar
incidence angle on the sunshield for DUNE will not significantly
perturb the payload performance, even for angles as large as 30
degrees.

\subsection{Visible FPA}

\begin{figure}
\begin{center}
\includegraphics[width=0.9\textwidth,angle=0]{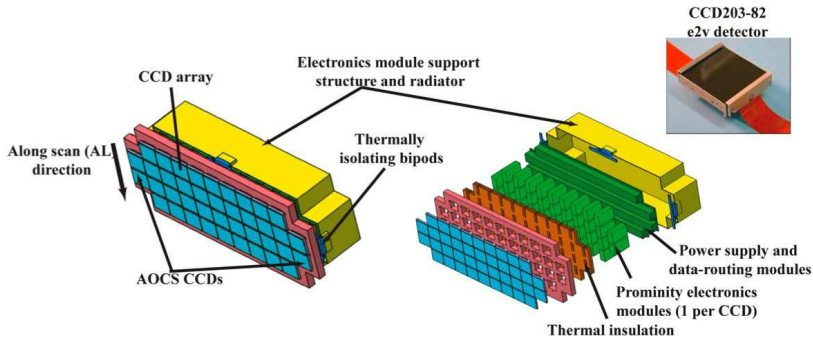}
\caption{Left: The VFP assembly with the 9x4 array of CCDs
and the 4 AOCS sensors on the front (blue) and the warm electronics
radiator at the back (yellow). Right: An expanded view of the VFP
assembly, including the electronics modules and thermal hardware (but
excluding the CCD radiator). Inset: The e2v CCD203-82 4kx4k pixels
shown here in a metal pack with flexi-leads for electrical
connections. One of the flexi-leads will be removed.  }
\label{fig:4.2}
\end{center}
\end{figure}

The visible Focal Plane Array (VFP) consists of 36 large format
red-sensitive CCDs, arranged in a 9x4 array (Figure~\ref{fig:4.2})
together with the associated mechanical support structure and
electronics processing chains. Four additional CCDs dedicated to
the AOCS measurements are located at the edge of
the array. All CCDs are 4096 pixel red-enhanced e2v CCD203-82 devices
with square 12 $\mu$m pixels.  The physical size of the array is
466x233 mm which corresponds to $1.09\deg \times 0.52 \deg$. Each pixel is
0.102 arcsec, so that the PSF is well sampled in each direction over
approximately 2.2 pixels, including all contributions.  The VFP
operates in the red band from 550-920nm. This bandpass is produced by
the dichroic.  The CCDs are 4-phase devices, so they can be clocked in
$1/4$ pixel steps. The exposure duration on each CCD is 375s,
permitting a slow readout rate and minimising readout noise. Combining
4 rows of CCDs will then provide a total exposure time of 1500s.

The VFP will be used by the spacecraft in a closed-loop system to
ensure that the scan rate and TDI clocking are synchronised. The two
pairs of AOCS CCDs provide two speed measurements on relatively bright
stars (V $\sim 22-23$). The DUNE VFP is largely a self-calibrating
instrument. For the shape measurements, stars of the appropriate
magnitude will allow the PSF to be monitored for each CCD including
the effects of optical distortion and detector alignment.
Radiation-induced charge transfer inefficiency will modify the PSF and
will also be self-calibrated in orbit.

\subsection{NIR FPA}

\begin{figure}
\begin{center}
\includegraphics[width=0.75\textwidth,angle=0]{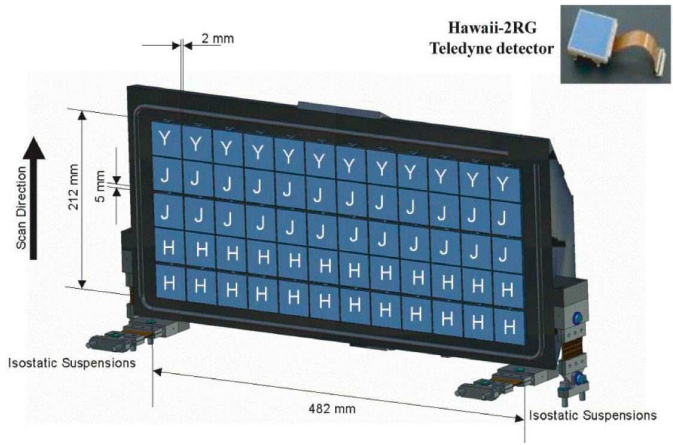}
\caption{Layout of the NIR FPA (MPE/Kayser-Threde). The 5
x 12 Hawaii 2RG Teledyne detector arrays (shown in the inset) are
installed in a molybdenum structure}
\label{fig:4.3}
\end{center}
\end{figure}

The NIR FPA consists of a 5 x 12 mosaic of 60 Hawaii 2RG detector
arrays from Teledyne, NIR bandpass filters for the wavelength bands Y,
J, and H, the mechanical support structure, and the detector readout
and signal processing electronics (see Figure~\ref{fig:4.3}). The FPA
is operated at a maximum temperature of 140 K for low dark current of
0.02$e^-$/s. Each array has 2048 x 2048 square pixels of 18 $\mu$m
size resulting in a 0.15 x 0.15 arcsec$^2$ field of view (FOV) per pixel.
The mosaic has a physical size of 482 x 212 mm, and covers a
FOV of $1.04^\circ \times 0.44^\circ$ or 0.46 square degrees.  The
HgCdTe Hawaii 2RG arrays are standard devices sensitive in the 0.8 to
1.7 $\mu$m wavelength range.

As the spacecraft is scanning the sky, the image motion on the NIR FPA
is stabilised by a de-scanning mirror during the integration time of
300s or less per NIR detector. The total integration time of 1500 s
for the $0.4^\circ$ high field is split among five rows and 3
wavelengths bands along the scan direction.  The effective integration
times are 600 s in J and H, and 300 s in Y. For each array, the
readout control, A/D conversion of the video output, and transfer of
the digital data via a serial link is handled by the SIDECAR ASIC
developed for JWST. To achieve the limiting magnitudes defined by the
science requirements within these integration times, a minimum of 13
reads are required.  Data are
processed in the dedicated unit located in the service module.
\section{Basic spacecraft key factors}
The spacecraft platform architecture is fully based on well-proven and
existing technologies. The mechanical, propulsion, and solar array
systems are reused from Venus Express (ESA) and Mars-Express. All the
AOCS, $\mu$-propulsion, Power control and DMS systems are reused from
GAIA. Finally, the science telemetry system is a direct reuse from the
PLEIADES (CNES) spacecraft. All TRLs are therefore high and all
technologies are either standard or being developed for GAIA (AOCS for
instance).

\subsection{Spacecraft architecture and configuration}
The spacecraft driving requirements are: (1) Passive cooling of both
visible and NIR focal planes below 170 K and 140 K, respectively; (2)
the PSF stability requirement, which translates to line of sight (LOS)
and payload thermal stability requirements; and (3) the high science
data rate. The spacecraft consists of a Payload Module (PLM) that
includes the instrument (telescope hardware, focal plane assemblies
and on board science data management) and a Service Module (SVM). The
SVM design is based on the Mars Express parallelepiped structure that
is 1.7 m $\times$ 1.7 m $\times$ 1.4 m, which accommodates all
subsystems (propulsion, AOCS, communication, power, sunshield, etc) as
well as the PLM. 
The spacecraft platform, and all technologies, are either
standard ones or being developed into GAIA programme (e.g. AOCS).

\subsection{Sunshield and Attitude Control}
The nominal scan strategy assumes a constant, normal ($0\deg$)
incidence of the sun on the sunshield, while allowing a sun incidence
angle of up to $30\deg$ to provide margin, flexibility for data
transmission manoeuvres and potential for further scan
optimisation. The sunshield is a ribbed, MLI-covered central frame
fixed to the platform.  The satellite rotates in a draft
scan-and-sweep-back approach where the spacecraft is brought back to
next scan position after each $20\deg$ strip. The scan rate is $1.12
\deg$ per hour, such that every day, one complete strip is scanned and
transmitted to ground.

Due to the observation strategy and the fixed high gain antenna (HGA),
the mission requires a high level of attitude manoeuvrability.
During data collection, the spacecraft is
rotated slowly about the sunshield axis. The slow scan control
requirements are equivalent to three-axis satellite control.  The
line-of-sight stability requirement is 0.2 $\mu$rad over 375s (the
integration time for one CCD) and is driven by optical quality and PSF
smearing, and will be partially achieved using
a continuous PSF calibration using the stars located in the
neighborhood (50 arcmin$^2$) of each observed galaxy. Detailed
analyses show that DUNE high pointing performance is comparable in
difficulty to that achieved on GAIA during science
observations. Similarly to GAIA, two pairs of dedicated CCD in the
visible focal plane are used for measuring the spacecraft attitude
speed vector. Hybridisation of the star tracker and payload
measurements is used to reduce the noise injected by the star tracker
in the loop. For all other operational phases and for the transition
from coarse manoeuvres to the science observation mode, the attitude
is controlled using the Mars Express propulsion system. The attitude
estimation is based on using two star trackers (also used in science
observing mode), six single-axis gyros and two sun sensors for
monitoring DUNE pointing during manoeuvres with a typically accuracy
better than 1 arcmin.

\subsection{Functional architecture: propulsion and electrical systems}
The star signal collected in the instrument is spread on the focal
plane assembly and transformed into a digital electrical signal which
is transferred to the Payload Data Handling Unit (PDHU), based on
Thales/AlienaSpace heritage. Power management and regulation are
performed by the Power Conditioning \& Distribution Unit (PCDU), and
based on the GAIA program. Electrical power is generated by two solar
arrays (2.7 m$^2$ each), as used in the Mars Express and Venus Express
ESA spacecraft. The control of their orientation is based on the
orientation of the whole spacecraft
towards the Sun. The panels are filled with AsGa cells.

The RF architecture is divided into two parts with the TT\&C system
(S-Band) plus a dedicated payload telemetry system (X-Band in the EES
band (Earth Exploration Service). The allocated bandwidth for payload
telemetry is 375 MHz and high rate transmitters already exist for
this purpose. The X-band 155 Mbits/s TMTHD modulator can be reused
from Pleiades spacecraft. A single fixed HGA of 30 cm diameter can be
used (re-used from Venus Express). The RF power required is 25 W, which
also enables the re-use of the solid state power amplifier (SSPA) from
Pleiades. The transmitted science data volume is estimated at 1.5
Tbits per day. The baseline approach consists in
storing the science data on board in the PDHU, then to downlink the
data twice per day. This can be achieved naturally twice per orbit at
06h and 18h local time and using the rotation degree of freedom about
the satellite-sun axis for orienting the antenna towards the ground
station. The total transmission duration is less than 3 hours. The
spacecraft attitude variation during transmission is less than 30 deg
(including AOCS margins). 20 kg hydrazine propellant budget is
required. In case the operational orbit would change to HEO, a dual
frequency (S-Band + X-Band) 35 m ESOC antenna could fit with the
mission needs, with in an increased HGA diameter (70 cm).

The required power on the GEO orbit is 1055 W. The
sizing case is the science mode after eclipse with battery
charging. Electrical power is generated by the two solar arrays of 2.7
m$^2$ each. With a $30\deg$ solar angle,
the solar array can generate up to 1150 W at the end of its life. The battery 
has been sized in a preliminary
approach for the eclipse case (64 Ah need).

\section{Science Operations and Data Processing}
The DUNE operational scenario follows the lines of a survey-type
project. The satellite will operate autonomously except for defined
ground contact periods during which housekeeping and science telemetry
will be downlinked, and the commands needed to control spacecraft and
payload will be uploaded.  The DUNE processing pipeline is inspired by
the Terapix pipeline used for the CFHT Legacy Survey.  The total
amount of science image data expected from DUNE is $\sim 370$
Terapixels (TPx): 150TPx from the Wide, 120TPx for 3 months of the
microlensing survey, 60TPx for the 3 months of the Galactic plane
survey, and 40TPx for 6 months deep survey. Based on previous
experience, we estimate an equal amount of calibration data (flat
fields, dark frames, etc.) will be taken over the course of the
mission. This corresponds to 740TB, roughly 20 times the amount of
science data for CFHT during 2003-2007.

There are four main activities necessary for the data processing,
handling, and data organisation of the DUNE surveys:
\begin{enumerate}
 \item software development: image and catalogue processing, 
 quality control, image and catalogue handling tools,
 pipeline development, survey monitoring, data archiving and
 distribution, numerical simulations, image simulations;
\item processing operation: running the pipeline, quality control and
quality assessment operation and versioning,
pipeline/software/database update and maintenance;
\item data archiving and data distribution: data and meta-data
products and product description, public user interface, external data
(non-DUNE) archiving and distribution, public outreach;
\item computing resources: data storage, cluster architecture,
GRID technology.
\end{enumerate}

\section{Conclusion: DUNE's Impact on Cosmology and Astrophysics}       

ESA's Planck mission will bring unprecedented precision to the
measurement of the high redshift Universe. This will leave the dark
energy dominated low redshift Universe as the next frontier in high
precision cosmology. Constraints from the radiation perturbation in
the high redshift CMB, probed by Planck, combined with density
perturbations at low redshifts, probed by DUNE, will form a complete
set for testing all sectors of the cosmological model. In this
respect, a DUNE+Planck programme can be seen as the next significant
step in testing, and thus challenging, the standard model of
cosmology. Table \ref{tableC2} illustrates just how precise the
constraints on theory are expected to be: DUNE will offer high
potential for ground-breaking discoveries of new physics, from dark
energy to dark matter, initial conditions and the law of gravity. Our
understanding of the Universe will be fundamentally altered in a
post-DUNE era, with ESA's science programmes at the forefront of these
discoveries.
As described above, the science goals of DUNE go far beyond the
measurement of dark energy. It is a mission which:
(i) measures both effects of dark energy (i.e. the expansion history
of the Universe and the growth of structure) by using weak lensing as the
central probe; (ii) places this high precision measurement of dark
energy within a broader framework of high precision cosmology by
constraining all sectors of the standard cosmology model (dark matter,
initial conditions and Einstein gravity); (iii) through a collection
of unique legacy surveys is able to push the frontiers of the
understanding of galaxy
evolution and the physics of the local group; and finally (iv) is able
to obtain information on some of the lowest masses astronomy
extrasolar planets, which could contain mirror Earths.

DUNE has been selected jointly with SPACE (Cimatti et al. 2008) in
ESA's Cosmic Vision programme for an assessment phase which lead to
the Euclid merged concept.

\begin{acknowledgements}
We thank CNES for support on an earlier version of the DUNE mission
and EADS/Astrium, Alcatel/Alenia Space, as well as Kayser/Threde for
their help in the preparation of the ESA proposal.

\end{acknowledgements}

\bibliographystyle{spbasic}      


\end{document}